\documentclass[12pt]{article}
\pdfoutput=1
\usepackage[margin=3cm]{geometry}
\usepackage[utf8]{inputenc}
\usepackage{amsmath}
\usepackage{amssymb}
\usepackage{authblk} 
\usepackage{xcolor}
\usepackage{physics}
\usepackage{graphicx} 
\usepackage{float}
\usepackage{tikz}
\usetikzlibrary{calc}
\usepackage{comment}

\usepackage{acronym}
\newacro{EFT}[EFT]{effective field theory}
\acrodefplural{EFT}{effective field theories}
\newacro{HO}[HO]{orthogonal}
\newacro{HE}[HE]{exceptional}
\newacro{NS5-brane}[NS5-brane]{Neveu-Schwarz five-brane}
\acrodefplural{NS5-brane}{Neveu-Schwarz five-branes}

\usepackage{hyperref}
\usepackage{varioref}    
\usepackage{cleveref}     

\crefname{table}{table}{tables}
\Crefname{table}{Table}{Tables}
\crefname{figure}{figure}{figures}
\Crefname{figure}{Figure}{Figures}

\definecolor{tealblue}{rgb}{0.21, 0.56, 0.63}
\definecolor{fadedSocialist}{RGB}{110, 47, 45}
\hypersetup{colorlinks=true,allcolors = tealblue,linktocpage=true}

\usepackage{pgf} 

\usepackage[
        backend=bibtex,
        sorting=none,
        style=phys,
        eprint=true,
        doi=false,
        biblabel=brackets
        ]{biblatex}

\bibliography{bibliography}

\usepackage[backgroundcolor=white!95!red, linecolor=cyan, bordercolor=cyan, textsize=footnotesize]{todonotes}

\newcommand{\commie}[1]{}

\numberwithin{equation}{section}
\numberwithin{table}{section}

\newenvironment{eqaed}
    {\begin{equation}
    \begin{aligned}
    }
    { 
    \end{aligned}
    \end{equation}
    \ignorespacesafterend
    }

\begin{document}

\title{\bf{Non-supersymmetric branes and discrete topological terms}}
\date{}

\author{Ivano Basile$^a$\thanks{\href{mailto:ibasile@mpp.mpg.de}{ibasile@mpp.mpg.de}} }
\author{Vittorio Larotonda$^{b,c}$\thanks{\href{mailto:vittorio.larotonda@unibo.it}{vittorio.larotonda@unibo.it}}}

\affil{${}^a$\emph{Max-Planck-Institut f\"ur Physik (Werner-Heisenberg-Institut)}\\ \emph{Boltzmannstraße 8, 85748 Garching, Germany}}

\affil{${}^b$\emph{Dipartimento di Fisica e Astronomia, Universit\`{a} di Bologna}\\ \emph{via Irnerio 46, Bologna, Italy}}
\affil{${}^c$\emph{INFN, Sezione di Bologna}\\ \emph{viale Berti Pichat 6/2, Bologna, Italy}}
\maketitle

\begin{abstract}

    In a recent work, Tachikawa and Zhang proved the existence of a discrete topological term in the unique non-supersymmetric heterotic string with no tachyons in ten dimensions. This theory features NS5-branes, whose chiral degrees of freedom are not well-understood due to the absence of dualities or supersymmetry. In this paper, we test the consistency of a tentative spectrum obtained by anomaly inflow, studying the relation between the worldvolume theory and the discrete topological term in spacetime. Furthermore, we conduct a bottom-up investigation of lower-dimensional gravitational theories with the same methods.
    
\end{abstract}

\thispagestyle{empty}

\newpage

\tableofcontents

\thispagestyle{empty}

\newpage

\pagenumbering{arabic}

\section{Introduction} \label{sec:introduction}

Topology and supersymmetry are amongst the most reliable tools to learn about non-perturbative physics, including quantum field theory and string theory. Topological methods are useful to probe global or discrete data of the theory, such as anomalies of the local \cite{Alvarez-Gaume:1983ict,Alvarez-Gaume:1983ihn, Alvarez-Gaume:1984zlq, Alvarez-Gaume:1984zst, Bilal:2008qx, Alvarez-Gaume:2022aak} and global \cite{Witten:1982fp, Witten:1985xe, Witten:2019bou} type, theta angles and, relatedly, the structure of the vacuum. Supersymmetry, when present, is a powerful organizational principle providing a computational framework for dynamically protected quantities.

For phenomenological as well as theoretical reasons, non-supersymmetric string theory recently garnered a renewed interest \cite{Abel:2016pwa, Mourad:2017rrl,  Abel:2018zyt, Basile:2018irz, Antonelli:2019nar, McGuigan:2019gdb, Itoyama:2020ifw, Angelantonj:2020pyr, Basile:2020mpt, Basile:2020xwi, Kaidi:2020jla, Faraggi:2020hpy, Cribiori:2020sct, Itoyama:2021fwc, Gonzalo:2021fma, Perez-Martinez:2021zjj, Basile:2021mkd, Basile:2021vxh, Sagnotti:2021mxb, Itoyama:2021itj, Mourad:2021roa, Cribiori:2021txm, Giri:2021eob, Raucci:2022bjw, Basile:2022ypo, Baykara:2022cwj, Koga:2022qch, Cervantes:2023wti, Angelantonj:2023egh, Avalos:2023mti, BoyleSmith:2023xkd, Raucci:2023xgx, Mourad:2023wjg, Mourad:2023ppi, Mourad:2023loc, Avalos:2023ldc, Fraiman:2023cpa, Basile:2023knk, DeFreitas:2024ztt, Tachikawa:2024ucm, Saxena:2024eil, Baykara:2024tjr, Baykara:2024vss, Angelantonj:2024jtu, Angelantonj:2024iwi, DeFreitas:2024yzr, Basaad:2024lno, Raucci:2024fnp, Abel:2024vov, Mourad:2024mpg, Larotonda:2024thv, Leone:2024hnr, Leone:2024xae, Basile:2025lek, Montero:2025ayi,Escalante-Notario:2025hvn,Hosseini:2025oka} since the discovery of several non-tachyonic settings \cite{Alvarez-Gaume:1986ghj, Dixon:1986iz, Sugimoto:1999tx}. While these perturbative constructions in Minkowski spacetime are free of tachyons, resulting in \textit{misaligned supersymmetry} \cite{Dienes:1995pm, Cribiori:2020sct, Cribiori:2021txm, Angelantonj:2023egh, Leone:2023qfd, Leone:2024xae}, their effective dynamics is plauged by runaway potentials associated to dilaton tadpoles \cite{Dudas:2000ff, Mourad:2017rrl}, whose consequences are largely unexplored, especially in lower dimensional settings \cite{Kaidi:2020jla, Fraiman:2023cpa, DeFreitas:2024ztt, Baykara:2024tjr, Angelantonj:2024jtu}. It is possible to build metastable weakly coupled vacua \cite{Basile:2018irz, Antonelli:2019nar, Baykara:2022cwj, Menet:2025nbf}, although their ultimate fate lies in the strongly coupled regime \cite{Basile:2022zee} due to brane nucleation. This state of affairs calls for a deeper understanding of the topological aspects of these theories. A step in this direction was taken in \cite{Basile:2023knk}\footnote{See also \cite{Hosseini:2025oka} for a treatment of the anomalies of such theories at the level of differential K-theory.}, where global anomalies of ten-dimensional non-supersymmetric strings were shown to cancel. These anomalies include changes in spacetime topology, as in the modern Dai-Freed viewpoint \cite{Garcia-Etxebarria:2018ajm}. Along these lines, fluctuations of spacetime topology in the context of holography and the swampland program led to the notion of cobordism triviality \cite{McNamara:2019rup, McNamara:2020uza}. This swampland requirement, deeply connected with the absence of global symmetries in gravity, spurred a number of investigations from the vantage point of string theory which also uncovered some novel non-supersymmetric objects \cite{Dierigl:2022reg, Debray:2023yrs, Dierigl:2023jdp, Braeger:2025kra}. The algebraic and topological techniques involved in the physics of cobordism defects goes hand in hand with the modern understanding of anomalies. In hindsight, this explains the connection between topological terms and anomalies at the global or discrete level.

This paper is mostly concerned with a particular discrete topological term in a non-supersymmetric heterotic string in ten dimensions, first discussed in \cite{Tachikawa:2024ucm}. This topological term is of gravitational origin and is valued in $\mathbb{Z}_3$, which means that it can in principle attain two non-zero values which could be detected by some observable. Since these terms cannot be captured by a local action, they can be defined by the assignment of a phase to certain manifolds via the (Euclidean) functional integral. This phase is a bordism invariant and takes the schematic form of a bilinear pairing between the theory and the spacetime manifold. The key insight of \cite{Tachikawa:2024ucm} is to relate this topological term to a global anomaly on the worldvolume of \acp{NS5-brane}. This can be viewed as an extension of the well-known relation between the Green-Schwarz mechanism \cite{Green:1984bx, Green:1984sg, Sagnotti:1992qw} and local anomaly inflow to global or discrete data. In \cite{Tachikawa:2024ucm}, the discrete topological term is shown not to vanish via a computation that involves the two supersymmetric heterotic strings in ten dimensions. This method avoids dealing with the problem of determining the chiral spectrum of non-supersymmetric \acp{NS5-brane}, which is difficult in the absence of string dualities.

However, anomaly inflow can provide at least some indications on plausible candidates for chiral \ac{NS5-brane} spectra \cite{Blaszczyk:2015zta, Basile:2023knk}. This prompts the question of whether a direct computation of the discrete topological term is possible, and whether the result agrees with that of \cite{Tachikawa:2024ucm}. In this paper we undertake this computation, largely following the method and notation of \cite{Tachikawa:2024ucm}. The contents of the paper are structured as follows. After a review of the topological term at stake in \cref{sec:discrete_top_term}, we retrace the computation of \cite{Tachikawa:2024ucm} in \cref{sec:global_anomaly_review}. In \cref{sec:non-susy_NS5} we discuss the proposal of \cite{Basile:2023knk} for the chiral spectrum of \acp{NS5-brane}, and we compute the resulting discrete topological term in \cref{sec:discrete_top_term_NS5}. We find the opposite value with respect to \cite{Tachikawa:2024ucm}, and discuss some possible explanations for the discrepancy. In \cref{sec:YM} we include spacetime Yang-Mills fields in the study of the \ac{NS5-brane} spectrum, showing that the proposal of \cite{Basile:2023knk} is consistent with the complete anomaly inflow. In \cref{sec:bottom-up} we extend our analysis to general gravitational effective theories from a bottom-up perspective. In particular, it turns out that non-supersymmetric chiral abelian gauge theories coupled to gravity feature discrete topological terms associated to the number of chiral fermions. We provide some concluding remarks in \cref{sec:conclusions}.

\section{Discrete topological term in heterotic strings}\label{sec:discrete_top_term}

In ten dimensions there are three perturbative heterotic vacua without tachyons \cite{Gross:1984dd,Gross:1985fr,Gross:1985rr, Alvarez-Gaume:1986ghj, Dixon:1986iz}. The two supersymmetric ones have an \ac{HO} and \ac{HE} gauge algebra, respectively $\mathfrak{so}(32)$ and $\mathfrak{e}_8 \oplus \mathfrak{e}_8$. The global structure of the gauge groups is not relevant for our discussion, but we shall denote them\footnote{The HE theory also features a semidirect $\mathbb{Z}_2$ factor \cite{McInnes:1999va}, which is relevant for CHL-like constructions \cite{Chaudhuri:1995bf, Chaudhuri:1995fk, Font:2021uyw, Nakajima:2023zsh, DeFreitas:2024ztt, Angelantonj:2024jtu} and anomalies \cite{Debray:2023rlx, Basile:2023knk}.}
\begin{eqaed}
    E_8 \times E_8 \, , \qquad \text{Spin}(32)/\mathbb{Z}_2 \, .
\end{eqaed}
The non-supersymmetric vacuum\footnote{Strictly speaking, a runaway potential is generated by string loops \cite{Alvarez-Gaume:1986ghj, Dixon:1986iz}. Nevertheless, metastable vacua can be found via flux compactification \cite{Basile:2018irz, Antonelli:2019nar, Baykara:2022cwj}.} has a $\mathfrak{so}(16) \oplus \mathfrak{so}(16)$ gauge algebra, and we shall denote it as $SO(16)^2$ abusing notation. The global structure of the gauge group is relevant for the considerations in \cite{McInnes:1999va, Basile:2023knk}.

While the field content and effective supergravity actions of the corresponding \acp{EFT} are well-understood \cite{deRoo:1992zp, Gross:1986mw}, there is room for discrete topological terms even in the absence of Yang-Mills fields. Indeed, turning off these fields, the gauge-invariant curvature $H$ of the $B$-field satisfies
\begin{eqaed}\label{eq:string_structure}
    dH = \frac{p_1}{2} \, ,
\end{eqaed}
where $p_1$ is (a representative of) the first Pontryagin class of the tangent bundle of spacetime, and $\frac{p_1}{2}$ denotes its canonical half on spin manifolds. When lifted to integral cohomology, this condition defines a \textit{string structure} \cite{Witten:1985mj, Sati:2009ic,Yonekura:2022reu} on spacetime.

Given a (stable) tangential structure $\xi$ on spacetime, the possible discrete topological terms in $d$ dimensions are classified by the torsion part of the bordism group $\Omega^\xi_d$ \cite{Freed:2004yc,Freed:2016rqq,Yonekura:2018ufj}. If $[M]\in\Omega^\xi_d$ is a nonzero torsion class of order $n$, one can introduce a $\mathbb{Z}_n$-valued topological term detecting the class $[M]$, which assigns the exponentiated Euclidean action $e^{2\pi i \frac{k}{n}}$ to the class $k[M]$. This type of topological term is well-defined when this phase is in fact a bordism invariant. For the two supersymmetric heterotic strings individually, this is not the case; however, the difference of the respective phases is a bordism invariant. As we shall recall in the following, this is also the well-defined discrete topological term for the non-supersymmetric heterotic string.

In the setting of interest, we have
\begin{eqaed}
    \Omega^{\text{string}}_{10} = \mathbb{Z}_6= \mathbb{Z}_2\times\mathbb{Z}_3 \, ,
\end{eqaed}
where the $\mathbb{Z}_3$ factor is generated by the group manifold $Sp(2)$ equipped with a unit of $H$-flux. In other words, the cohomology class of $H$ is the generator $1 \in H^3(Sp(2),\mathbb{Z}) = \mathbb{Z}$. In order to generate the $\mathbb{Z}_2$ factor one has to introduce exotic spheres; for our purposes it is sufficient to focus on the $\mathbb{Z}_3$ subgroup, following \cite{Tachikawa:2024ucm} -- this restriction will be justified in \cref{sec:bottom-up}. This means that any ten-dimensional closed string manifold that we consider is cobordant to either zero, one or two copies of this fluxed $Sp(2)$ manifold.

Heterotic string theory also contains non-abelian Yang-Mills fields and \acp{NS5-brane}, which can be seen as the zero-size limit of their solitonic configuration \cite{Strominger1990, Callan:1991ky,Kutasov1995, Garcia-Etxebarria:2014txa}. For this reason, the $Sp(2)$ generator does bound an eleven-dimensional manifold $N_{11}$ if we allow a non-zero gauge field $F$ on $N_{11}$ such that
\begin{eqaed}
    dH = \frac{p_1 + c_2(F)}{2} \, .
\end{eqaed}
To see this, we recall that $Sp(2)$ is an $S^3$-fibration over $S^7$, and $S^3$ with an unit of $H$-flux is the boundary of a disc $D^4$ with one unit of instanton number. The disk fibration of $D^4$ over $S^7$ provides the bounding manifold.

In \cite{Tachikawa:2023lwf} the topological term associated to the $Sp(2)$ generator is computed. The only necessary data are the Green-Schwarz terms
\begin{eqaed}
    B_2 \wedge Y^{E_8\times E_8} \, , \qquad B_2 \wedge Y^{\text{Spin}(32)/\mathbb{Z}_2} \, .
\end{eqaed}
Furthermore, the method of \cite{Tachikawa:2024ucm} can be used to compute the analogous term in the non-supersymmetric $SO(16)^2$ string, since
\begin{eqaed}\label{eq:X_8 subtraction}
    Y^{SO(16)^2} = Y^{\text{Spin}(32)/\mathbb{Z}_2} - Y^{E_8\times E_8} \, .
\end{eqaed}
In the following we illustrate the method to compute the discrete topological term via anomaly inflow onto \acp{NS5-brane} in these three settings. Our conventions are the following:
\begin{itemize}
    \item $Q$ is the tangent bundle of the ten-dimensional spacetime;
    \item $T$ is the tangent bundle to the worldvolume of the \ac{NS5-brane};
    \item $N$ is the normal bundle to the worldvolume of the \ac{NS5-brane}.
\end{itemize}
As vector bundles
\begin{eqaed}
    Q|_\text{worldvolume} = T \oplus N \, ,
\end{eqaed}
while the first Pontryagin class splits according to
\begin{eqaed}
    p_1(Q) = p_1(T) + p_1(N) \, .
\end{eqaed}
The residual rotational symmetry of $N$ decomposes as $SU(2)_L \times SU(2)_{\text{R}}$, where $SU(2)_{\text{R}}$ is the R-symmetry of the $\mathcal{N}=(0,1)$ supersymmetry preserved by the \acp{NS5-brane}. In the non-supersymmetric theory this interpretation is lost, but we shall nevertheless focus on the $SU(2)_{\text{R}}$ subgroup. We normalize field strengths $F$ to include factors of $2\pi$, so that the Chern-Weyl expressions for characteristic classes take the form
\begin{eqaed}
    & p_1(F) = -\frac{1}{2} \tr F^2 \, , \\
    & p_2(F) = \frac{1}{8} (\tr F^2)^2 - \frac{1}{4}\tr F^4 = \frac{(\tr F^2)^2 - 2\tr F^4}{8} \, .
\end{eqaed}
Furthermore, in terms of the $SU(2)_{\text{L}}$ and $SU(2)_{\text{R}}$ Chern classes, the Pontryagin and Euler classes of the normal bundle can be written as
\begin{eqaed}
    & \frac{p_1(N)}{2} = -\, c_2(L) - c_2(R) \, , \\
    & \chi (N) = c_2(L) - c_2(R) \, .
\end{eqaed}

\subsection{The local Green-Schwarz mechanism}\label{sec:GS_mechanism}

The anomaly polynomials of all heterotic strings factorize compatibly with the Green-Schwarz mechanism for local anomaly cancellation \cite{Green:1984sg, Schellekens:1986xh, Lerche:1987sg, Sagnotti:1992qw}. The $\text{Spin}(32)/\mathbb{Z}_2$ string has
\begin{eqaed}\label{eq:Anomaly Polynomial Spin(32)}
    I_{12}^{\text{Spin}(32)/\mathbb{Z}_2} = X_4^{\text{Spin}(32)/\mathbb{Z}_2} \wedge Y_8^{\text{Spin}(32)/\mathbb{Z}_2} \, ,
\end{eqaed}
with
\begin{eqaed}
    & X_4^{\text{Spin}(32)/\mathbb{Z}_2}= \frac{p_1(Q)}{2}-\frac{p_1(F)}{2} \, , \\
    & Y_8^{\text{Spin}(32)/\mathbb{Z}_2}=\frac{3 p_1^2(Q)-4p_2(Q)}{192}-\frac{p_1(Q)p_1(F)}{48}+ \frac{p_1^2(F)-2p_2(F)}{12} \, .
\end{eqaed}
Note that, compared to \cite{Mourad:1997uc}, there is an overall minus sign, coming from an opposite convention on the chirality.

For the $E_8\times E_8$ heterotic theory, we similarly have
\begin{eqaed}
    I_{12}^{E_8\times E_8} = X_4^{E_8\times E_8} \wedge Y_8^{E_8\times E_8} \, .
    \label{eq:Anomaly Polynomial E8xE8}
\end{eqaed}
In this case, denoting the two $E_8$ field strengths by $F_1$ and $F_2$,
\begin{eqaed}
    X_4^{E_8\times E_8} &= \frac{p_1(Q)}{2}-\frac{p_1(F_1)}{2}-\frac{p_1(F_2)}{2}\,,\\
     Y_8^{E_8\times E_8}&=\frac{3 p_1^2(Q)-4p_2(Q)}{192}-\frac{p_1(Q)(p_1(F_1)+p_1(F_2))}{48} \\
     & + \frac{p_1^2(F_1)}{12} - \frac{p_1(F_1)p_1(F_2)}{24} +\frac{p_1^2(F_2)}{12}\,.
\end{eqaed}
In both cases the anomaly is cancelled via the Green-Schwarz mechanism, which involves a 2-form gauge field $B_2$ whose (gauge-invariant) field strength $H$ satisfies
\begin{eqaed}
    dH = X_4\,.
\end{eqaed}
The effective action contains the Green-Schwarz term
\begin{eqaed}\label{eq:GS}
    -\int B_2 \wedge Y_8 \, ,
\end{eqaed}
whose gauge variation cancels the gauge anomaly encoded in \eqref{eq:Anomaly Polynomial Spin(32)} and \eqref{eq:Anomaly Polynomial E8xE8} by Wess-Zumino descent.

\subsection{Anomaly inflow onto heterotic branes}\label{sec:anomaly_inflow_review}

The class $Y_8$ in the Green-Schwarz term also plays a role in the anomaly of the worldvolume theory living on \acp{NS5-brane}. In particular, the anomaly polynomial of this theory contains $Y_8$, as we shall recall in more detail shortly. These two quantities are connected by anomaly inflow. This is the key mechanism to compute the discrete topological term of \cite{Tachikawa:2024ucm}: the local worldvolume anomaly is connected to the local Green-Schwarz term of \cref{eq:GS} as the global worldvolume anomaly is connected to the discrete topological term, which is a global version of \cref{eq:GS}.

To see the connection at the local level, recall that the Green-Schwarz coupling can be locally recast into a Chern-Simons coupling,
\begin{equation}
    \int_{M_{10}} B_2 \wedge Y_8 = \int_{M_{10}} H_2 \wedge \text{CS}(Y_8) \, .
    \label{eq:GS term}
\end{equation}
We consider the ten dimensional spacetime $M_{10}$ to be a spherical fibration of the form
\begin{equation}
    S^3 \hookrightarrow M_{10} \rightarrow N_7\,,
\end{equation}
with
\begin{equation}
    \int_{S^3}H = k \, .
\end{equation}
Then, integrating over fibers yields
\begin{equation}
    k \int_{N_7} \text{CS}(Y_8) \, ,
\end{equation}
namely the anomaly inflow onto the \ac{NS5-brane}. Indeed, if $N_7$ describes the worldvolume of a \ac{NS5-brane} together with its radial transverse direction, the $S^3$ fiber shrinks to zero radius at the location of the brane, producing the same gauge variation of the action as the one arising from a Dirac delta current localized on the worldvolume. In particular, this procedure captures the anomaly (inflow) of the rotational symmetry $SU(2)_{\text{R}}$ of the normal bundle. Once the global anomaly is included, this is sufficient to detect the $\mathbb{Z}_3$-valued discrete topological term. The strategy is outlined in \cref{fig:diagram}.

\tikzset{every picture/.style={line width=0.75pt}} 

\begin{figure}
    \centering     

\begin{tikzpicture}[x=0.75pt,y=0.75pt,yscale=-1,xscale=0.9]

\draw  [color={rgb, 255:red, 163; green, 9; blue, 29 }  ,draw opacity=1 ][fill={rgb, 255:red, 219; green, 81; blue, 21 }  ,fill opacity=0 ][line width=3]  (32,73.31) .. controls (32,67.89) and (36.4,63.49) .. (41.83,63.49) -- (204.61,63.49) .. controls (210.04,63.49) and (214.44,67.89) .. (214.44,73.31) -- (214.44,102.79) .. controls (214.44,108.22) and (210.04,112.62) .. (204.61,112.62) -- (41.83,112.62) .. controls (36.4,112.62) and (32,108.22) .. (32,102.79) -- cycle ;
\draw  [color={rgb, 255:red, 48; green, 88; blue, 5 }  ,draw opacity=1 ][fill={rgb, 255:red, 184; green, 233; blue, 134 }  ,fill opacity=0 ][line width=3]  (445,76.31) .. controls (445,70.89) and (449.4,66.49) .. (454.83,66.49) -- (617.61,66.49) .. controls (623.04,66.49) and (627.44,70.89) .. (627.44,76.31) -- (627.44,105.79) .. controls (627.44,111.22) and (623.04,115.62) .. (617.61,115.62) -- (454.83,115.62) .. controls (449.4,115.62) and (445,111.22) .. (445,105.79) -- cycle ;
\draw    (217.04,88.85) -- (439.24,89.8) ;
\draw [shift={(442.24,89.82)}, rotate = 180.25] [fill={rgb, 255:red, 0; green, 0; blue, 0 }  ][line width=0.08]  [draw opacity=0] (8.93,-4.29) -- (0,0) -- (8.93,4.29) -- cycle    ;
\draw  [color={rgb, 255:red, 48; green, 88; blue, 5 }  ,draw opacity=1 ][fill={rgb, 255:red, 184; green, 233; blue, 134 }  ,fill opacity=0 ][line width=3]  (445,214.31) .. controls (445,208.89) and (449.4,204.49) .. (454.83,204.49) -- (617.61,204.49) .. controls (623.04,204.49) and (627.44,208.89) .. (627.44,214.31) -- (627.44,243.79) .. controls (627.44,249.22) and (623.04,253.62) .. (617.61,253.62) -- (454.83,253.62) .. controls (449.4,253.62) and (445,249.22) .. (445,243.79) -- cycle ;
\draw   (517,139.41) -- (534.9,118.91) -- (552.79,139.41) -- (543.85,139.41) -- (543.85,180.41) -- (552.79,180.41) -- (534.9,200.91) -- (517,180.41) -- (525.95,180.41) -- (525.95,139.41) -- cycle ;
\draw  [color={rgb, 255:red, 170; green, 156; blue, 12 }  ,draw opacity=1 ][fill={rgb, 255:red, 184; green, 233; blue, 134 }  ,fill opacity=0 ][line width=3]  (263.23,258.31) .. controls (263.23,252.89) and (267.63,248.49) .. (273.05,248.49) -- (425.61,248.49) .. controls (431.04,248.49) and (435.44,252.89) .. (435.44,258.31) -- (435.44,287.79) .. controls (435.44,293.22) and (431.04,297.62) .. (425.61,297.62) -- (273.05,297.62) .. controls (267.63,297.62) and (263.23,293.22) .. (263.23,287.79) -- cycle ;
\draw  [color={rgb, 255:red, 170; green, 156; blue, 12 }  ,draw opacity=1 ][fill={rgb, 255:red, 184; green, 233; blue, 134 }  ,fill opacity=0 ][line width=3]  (262.23,172.31) .. controls (262.23,166.89) and (266.63,162.49) .. (272.05,162.49) -- (424.61,162.49) .. controls (430.04,162.49) and (434.44,166.89) .. (434.44,172.31) -- (434.44,201.79) .. controls (434.44,207.22) and (430.04,211.62) .. (424.61,211.62) -- (272.05,211.62) .. controls (266.63,211.62) and (262.23,207.22) .. (262.23,201.79) -- cycle ;
\draw    (476,256) .. controls (476,270.23) and (469.41,276.3) .. (439.81,276.65) ;
\draw [shift={(437,276.67)}, rotate = 360] [fill={rgb, 255:red, 0; green, 0; blue, 0 }  ][line width=0.08]  [draw opacity=0] (8.93,-4.29) -- (0,0) -- (8.93,4.29) -- cycle    ;
\draw    (476.44,201.32) .. controls (476.44,183.95) and (464.23,184.91) .. (439.12,185.89) ;
\draw [shift={(436.33,186)}, rotate = 357.88] [fill={rgb, 255:red, 0; green, 0; blue, 0 }  ][line width=0.08]  [draw opacity=0] (8.93,-4.29) -- (0,0) -- (8.93,4.29) -- cycle    ;
\draw    (258.1,187.36) .. controls (190.13,188.35) and (152.24,176.72) .. (151.12,118.07) ;
\draw [shift={(151.1,115.36)}, rotate = 90] [fill={rgb, 255:red, 0; green, 0; blue, 0 }  ][line width=0.08]  [draw opacity=0] (8.93,-4.29) -- (0,0) -- (8.93,4.29) -- cycle    ;
\draw    (258.28,268.28) .. controls (142.86,267.28) and (76.94,231.64) .. (75.12,117.1) ;
\draw [shift={(75.1,115.36)}, rotate = 89.41] [fill={rgb, 255:red, 0; green, 0; blue, 0 }  ][line width=0.08]  [draw opacity=0] (8.93,-4.29) -- (0,0) -- (8.93,4.29) -- cycle    ;

\draw (70,25) node [anchor=north west][inner sep=0.75pt]   [align=left] {{\large \textbf{Spacetime}}};
\draw (465,28) node [anchor=north west][inner sep=0.75pt]   [align=left] {{\large \textbf{Worldvolume}}};
\draw (45,80) node [anchor=north west][inner sep=0.75pt]   [align=left] {Green-Schwarz term};
\draw (475,83) node [anchor=north west][inner sep=0.75pt]   [align=left] {Anomaly inflow};
\draw (486,220) node [anchor=north west][inner sep=0.75pt]   [align=left] {NS5 anomaly};
\draw (554,147) node [anchor=north west][inner sep=0.75pt]   [align=left] {\textbf{Matching}};
\draw (290,179) node [anchor=north west][inner sep=0.75pt]   [align=left] {Global anomaly};
\draw (290,264) node [anchor=north west][inner sep=0.75pt]   [align=left] {Local anomaly};
\draw (168,144) node [anchor=north west][inner sep=0.75pt]   [align=left] {Discrete};
\draw (126,214) node [anchor=north west][inner sep=0.75pt]   [align=left] {Continuous};

\end{tikzpicture}

    \caption{A diagram of how the discrete and continuous topological terms of the spacetime theory are encoded in the global and local anomalies of the worldvolume theory of \acp{NS5-brane} via inflow.}
    \label{fig:diagram}
\end{figure}
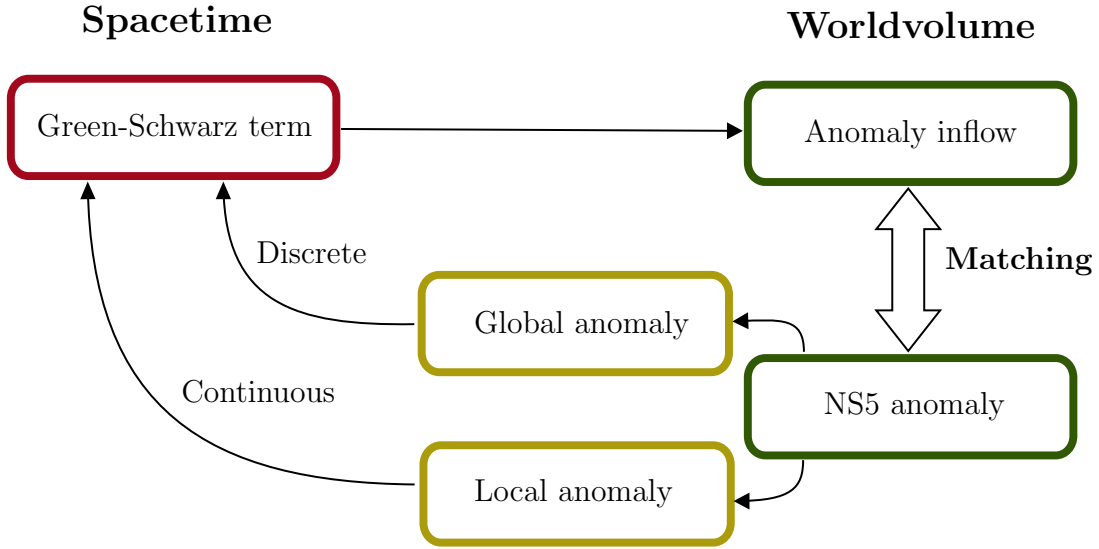

The $\mathbb{Z}_3$-valued is detected by the $Sp(2)$ generator with one unit of flux. The spherical fibration is defined by the inclusion $Sp(1) \subset Sp(2)$ given by
\begin{equation}
     q \in Sp(1) \mapsto \text{diag}(q,1)\in Sp(2) \, .
\end{equation}
Since $Sp(1) \cong SU(2) \cong S^3$, there is a fibration
\begin{equation}
    S^3 \cong Sp(1) \hookrightarrow Sp(2) \rightarrow Sp(2)/Sp(1) \cong S^7\,,
\end{equation}
where $H$ is supported on the fiber.
Bundles of this type are classified up to homotopy by homotopy classes of maps to the classifying space $SU(2)$, namely
\begin{equation}
    [S^7 \to BSU(3)] = \pi_7(BSU(2)) = \pi_6(SU(2)) = \mathbb{Z}_{12}\,,
\end{equation}
and this particular fibration corresponds to the generator. The phase of partition function of the spacetime theory evaluated on this manifold measures the global anomaly of the worldvolume theory of \acp{NS5-brane} for the normal bundle rotational symmetry, as explicitly shown in \cite{Tachikawa:2024ucm} for the $SU(2)_{\text{R}}$ subgroup.

\subsubsection*{NS5-branes in the HO
 heterotic string}
 
The \ac{NS5-brane} in $\text{Spin}(32)/\mathbb{Z}_2$ heterotic theory is dual to the D5-brane in Type I string theory \cite{Polchinski:1995df}. Therefore, the worldvolume theory can be obtained by quantizing open strings, and anomaly cancellation via inflow has been derived in \cite{Dixon:1992if, Mourad:1997uc, Imazato:2010qz}.
On the worldvolume of a stack of $k$ NS5-branes, the theory consists of a $\mathcal{N}=(0,1)$ supersymmetric gauge theory with an $\mathfrak{sp}(k)$ gauge algebra \cite{Witten:1995gx, Schwarz:1995zw} coupled to an hypermultiplet in the antisymmetric representation of $\mathfrak{sp}(k)$ and an half-hypermultiplet in the bifundamental representation of $\mathfrak{sp}(k)\oplus\mathfrak{so}(32)$. The gauginos are also doublets of $SU(2)_{\text{R}}$, the antisymmetric hypermultiplets are doublets of $SU(2)_{\text{L}}$ and the bifundamental ones are neutral. Computing the chiral anomaly polynomial $I_8^{Sp(k)}$ of the six-dimensional theory and subtracting the anomaly inflow term $k \, Y_8$, one has
\begin{equation}
    I_8^{Sp(k)} - k Y_8^{\text{Spin}(32)/\mathbb{Z}_2} = Z_4^{\text{Spin}(32)/\mathbb{Z}_2} \wedge W_4^{\text{Spin}(32)/\mathbb{Z}_2} \, ,
    \label{eq:Spin(32) inflow}
\end{equation}
where
\begin{eqaed}
    Z_4^{\text{Spin}(32)/\mathbb{Z}_2} & = X_4^{\text{Spin}(32)/\mathbb{Z}_2} + k \, \chi(N)\,, \\
    W_4^{\text{Spin}(32)/\mathbb{Z}_2} & = k \, \frac{p_1(T) - p_1(N)}{24}-p_1(G)\,.
\end{eqaed}
Here $G$ is the worldvolume field strength of $\mathfrak{sp}(k)$. In presence of a stack of \acp{NS5-brane}, the $H$-field localized on the worldvolume satisfies
\begin{equation}
    dH= Z_4^{\text{Spin}(32)/\mathbb{Z}_2} =\frac{p_1(Q)}{2}-\frac{p_1(F)}{2} + k \, \chi(N) \, .
\end{equation}
This comes from the decomposition $H_\text{tot} = H + H_\text{fiber}$, where the fiber contribution satisfies $dH_\text{fiber}= - k \, \chi(N)$. This additional term in the anomaly polynomial leads to an additional Green-Schwarz term on the worldvolume \cite{Mourad:1997uc}, which reads
\begin{equation}
    -\int B_2 \wedge  W_4^{\text{Spin}(32)/\mathbb{Z}_2} \, .
\end{equation}
This is responsible of the cancellation of the remnants of the subtraction between the chiral anomaly and the inflow in \eqref{eq:Spin(32) inflow}. In particular, as pointed out in \cite{Tachikawa:2024ucm}, this term cancels the mixed gauge anomaly on the worldvolume. Assuming that $\mathfrak{sp}(k)$ is free of global anomalies as well, we can turn off the corresponding gauge field consistently and focus on the normal bundle anomaly.

\subsubsection*{NS5-branes in the HE
 heterotic string}
 
In the HE case, a stack of $k$ \acp{NS5-brane} host a rank-$k$ E-string theory \cite{Ganor:1996mu, Seiberg:1996vs}. The anomaly polynomial of such a theory has been computed in \cite{Ohmori:2014pca} via inflow from M-theory. Letting the E-string theory correspond to the first $E_8$ factor,
\begin{equation}
    I_8^{\text{rank-}k} - k Y_8^{E_8\times E_8} = Z_4^{E_8\times E_8} \wedge W_4^{E_8\times E_8} \, ,
\label{eq:E8xE8 inflow}
\end{equation}
with
\begin{equation}
    Z_4^{E_8\times E_8} = X_4^{E_8\times E_8} + k \, \chi(N)\,,
\end{equation}
and
\begin{equation}
    W_4^{E_8 \times E_8} = k \, \frac{p_1(T) + p_1(N)}{24}+\frac{\chi(N) - p_1(F_1)}{6} + \frac{p_1(F_2)}{12} \, .
\end{equation}
In this case the $H$-field satisfies
\begin{equation}
    dH = Z_4^{E_8\times E_8} = \frac{p_1(Q)-p_1(F_1)-p_1(F_2)}{2}+k\chi(N) \, .
\end{equation}
Once again, the worldvolume hosts a localized Green-Schwarz term
\begin{equation}
    -\int B_2 \wedge W_4^{E_8\times E_8} \, .
\end{equation}

\subsubsection*{Difference of the two cases}

As originally discussed in \cite{Alvarez-Gaume:1986ghj, Schellekens:1986xh}, and recently revisited in \cite{Basile:2023knk, Tachikawa:2024ucm}, the virtual difference of the HE and HO chiral spectra, restricted to the common $\mathfrak{so}(16) \oplus \mathfrak{so}(16)$ subalgebra, yields the chiral content of the non-supersymmetric heterotic string. Therefore, the difference of the (restricted) anomaly polynomials follows the same pattern. Taking $k=1$ for simplicity, and onlt turning on the $SU(2)_{\text{R}}$ bundle, for both HE and HO theories one obtains
\begin{eqaed}
    X_4 & = \frac{p_1(Q)}{2} \, , \\
    Y_8 & = \frac{3p_1^2(Q) - 4p_2(Q)}{192} \, , \\
    Z_4 & = \frac{p_1(T)}{2} - c_2(R) \, .
\end{eqaed}
The respective worldvolume anomaly polynomials then simplify to
\begin{eqaed}
     & I_8^{Sp(k)} - k Y_8^{\text{Spin}(32)/\mathbb{Z}_2} = Z_4 \wedge \frac{p_1(T)-p_1(N)}{24} \, , \\
     & I_8^{\text{rank-}k} - k Y_8^{E_8\times E_8} = Z_4 \wedge \left(\frac{p_1(T)+p_1(N)}{24}-\frac{c_2(R)}{6}\right) .
\end{eqaed}
Using $p_1(N) = -2c_2(R)$, the difference is
\begin{equation}\label{eq:virtual_difference}
    I_8^{Sp(k)} - I_8^{\text{rank-}k} = Z_4 \wedge \left(-\frac{p_1(N)}{12}+\frac{c_2(R)}{6}\right)=Z_4 \wedge \frac{c_2(R)}{3}  \, ,
\end{equation}
corresponding to the non-supersymmetric worldvolume Green-Schwarz coupling
\begin{equation}
    -\int B_2 \wedge \frac{c_2(R)}{3} \,.
\end{equation}

\subsection{Computation of the global anomaly}\label{sec:global_anomaly_review}
We now turn to the global worldvolume anomaly. On the one hand, the Witten anomaly \cite{Witten:1982fp} for $SU(2)_{\text{R}}$ is classified by \cite{Kiritsis:1986mf, Tosa:1989qm, Bershadsky:1997sb, Lee:2020ewl,Davighi:2020kok}
\begin{equation}
    \pi_6(SU(2))=\mathbb{Z}_{12} \, .
\end{equation}
The global anomaly arises from the Green-Schwarz term \cite{Tosa:1989qm, Bershadsky:1997sb, Lee:2020ewl,Davighi:2020kok, Tachikawa:2024ucm}, which requires that the worldvolume and the seven-dimensional anomaly background $N_7$ be endowed with a \textit{twisted string structure}\footnote{In the context of string theory, the impact of the twisted string structure on bordism groups was discussed in \cite{Dierigl:2022zll, Basile:2023knk, Basile:2023zng, Kneissl:2024zox}. The cancellation of global anomalies in supersymmetric heterotic strings also follows from the Stolz-Teichner conjecture on topological modular forms \cite{Tachikawa:2021mby, Tachikawa:2021mvw}.} defined by the Bianchi identity $dH = Z_4$. Indeed, on mere spin manifolds, six-dimensional $SU(2)$ gauge theories devoid of local anomalies have no global anomalies either, since the relevant bordism group $\Omega^{\text{spin}}_7(BSU(2)) = 0$. In order to simplify the ensuing discussion, we denote $p_1\equiv p_1(T)$ and $c_2\equiv c_2(R)$.

\subsubsection*{Strategy}

Taking stock of our preceding considerations, the discrete topological term for the non-supersymmetric heterotic string is well-defined and is captured by the global anomaly of the corresponding \ac{NS5-brane}. In \cite{Tachikawa:2024ucm}, this anomaly is computed from the virtual difference in \cref{eq:virtual_difference}. Our aim is to compute this topological term directly in the non-supersymmetric theory, in terms of the chiral spectrum proposed in \cite{Basile:2023knk}. We begin by recalling the method employed in \cite{Tachikawa:2024ucm} along the lines of \cite{Kiritsis:1986mf, Tosa:1989qm, Bershadsky:1997sb, Lee:2020ewl}. Consider a $SU(2)$ theory whose anomaly polynomial reads
\begin{equation}
    I_8 = Z_4 \wedge W_4 \, ,
\end{equation}
endowed with a twisted string structure defined by the Bianchi identity $dH = Z_4$. The global anomaly is described by the seven-dimensional anomaly theory
\begin{equation}
    X \equiv \eta(N_7)-\int_{N_7} H_3\wedge W_4
    \label{eq:7d anomaly}
\end{equation}
intended modulo unity, i.e. as a normalized phase or equivalently an element of $\mathbb{R}/\mathbb{Z}$. The first term is the eta invariant of the Dirac operator twisted by the virtual representation of the chiral fermions in the theory. Since $\Omega_7^{\text{spin}}(BSU(2))=0$, this invariant can be written as the integral of $I_8$ on any eight-dimensional extension of $N_7$ due to the Atiyah-Patodi-Singer theorem, and thus only depends on $I_8$. The second term lifts the worldvolume Green-Schwarz term. If the anomaly background $N_7$ is nullbordant in twisted string bordism, $H$ can be extended to eight dimensions too, yielding $X=0$. This is a straightforward way of seeing that the obstruction to anomaly cancellation lies in the twisted string structure. Turning the logic around, the results of \cite{Tachikawa:2024ucm} show that the relevant twisted string bordism does not vanish.

In the following we choose $N_7 = S^7$, which probes the Witten anomaly. More general manifolds that are non-trivial in twisted string bordism would probe Dai-Freed anomalies \cite{Garcia-Etxebarria:2018ajm}. As we have derived in the preceding section, the anomaly polynomial $I_8$ obtained subtracting the two supersymmetric spectra is given by
\begin{eqaed}\label{eq:ZW_het}
    Z_4 & =\frac{p_1}{2}-2c_2 \, , \\
    W_4 & =\frac{c_2}{3} \, .
\end{eqaed}
Since the local anomaly cancels, the anomaly theory on $S^7$ defines a homomorphism
\begin{equation}
    \mathcal{A}:\pi_6(SU(2))\longrightarrow \mathbb{R}/\mathbb{Z}\,.
\end{equation}
Indeed, since $H^3(S^7,\mathbb{R})=0$ any two solutions $H$ and $H'$ of the Bianchi identity are related by $H=H'+dB$ for some globally defined $B$. Hence, the anomaly does not depend on this choice, and only depends on the $SU(2)$ bundle. Using this fact, together with the long exact sequence of homotopy groups associated to the spherical fibration of $Sp(2)$, the global anomaly on the generator of $\pi_6(SU(2))$ can be evaluated via anomaly interplay viewing $SU(2) \cong Sp(1)$ as a subgroup of $Sp(2)$. The latter has no Witten anomaly, and thus the $SU(2)$ global anomaly is given by the $Sp(2)$ global anomaly \cite{Kiritsis:1986mf, Tosa:1989qm, Bershadsky:1997sb, Lee:2020ewl, Tachikawa:2024ucm}.

Choosing a virtual $Sp(2)$ representation whose restriction to $SU(2)$ yields the anomaly polynomial $I_8$, one ends up with the integral of the $Sp(2)$ anomaly polynomial on the eight-disk bounding $S^7$. This equals to the integral over $S^8$ obtained gluing the disk to another disk\footnote{More precisely, this is done for the trivial class in $\pi_6(SU(2))$ which is twelve times the generator, and finally dividing by twelve \cite{Tachikawa:2024ucm}.}. The result is independent on the choice of $Sp(2)$ representation extending the $SU(2)$ chiral spectrum. Concretely, any such extension has an anomaly polynomial of the form
\begin{equation}
    \tilde{I}_8 = \frac{t}{48}\text{tr}F^4 +J_8(c_2,p_1) \, ,
\end{equation}
which reduces to $Z_4 \wedge W_4$ via the branching rule
\begin{equation}
    \text{tr}F^4 \to \frac{1}{2}(\text{tr}F^2)^2 = 2(c_2)^2 \, .
\end{equation}
Then, the anomaly on the generator of $\pi_6(SU(2))$ is \cite{Tachikawa:2024ucm}
\begin{equation}
    \mathcal{A}(1) =\frac{t}{12} \, .
\end{equation}

\subsubsection{Evaluation}

Here we reproduce the computation of $t$ in \cite{Tachikawa:2024ucm} for the reader's convenience, setting up the branching rules and conventions that we shall use in the following sections as well. The goal is to find an $Sp(2)$ system whose anomaly polynomial matches \cref{eq:ZW_het}
when restricted to the $SU(2)\cong Sp(1)$ subgroup of $Sp(2)$.

First, we express the total anomaly polynomial as a linear combination of the anomaly polynomials of the \textbf{1}, \textbf{2} and \textbf{3} representations of $SU(2)$. The corresponding contributions are
\begin{eqaed}
    I_\textbf{1} & = \frac{7p_1^2 - 4p_2}{5760} \, , \\
    I_\textbf{2} & = \frac{1}{24}p_1c_2 +\frac{1}{12}c_2^2+ 2I_\textbf{1} \, \\
    I_\textbf{3} & = \frac{1}{6}p_1c_2 +\frac{4}{3}c_2^2+ 3 I_\textbf{1} \, .
\end{eqaed}
Imposing that a generic linear combination
\begin{equation}
    \alpha I_\textbf{2} +\beta I_\textbf{3} + \gamma I_\textbf{1}
\end{equation}
match $Z_4 \wedge W_4$, we find
\begin{eqaed}
    \begin{cases}
         \frac{1}{24}\alpha  +\frac{1}{6}\beta = &\frac{1}{6} \, , \\
         \frac{1}{12}\alpha + \frac{4}{3}\beta= & -\frac{2}{3} \, ,
    \end{cases}
\end{eqaed}
whose solution is
\begin{equation}
    \alpha = 8 \, , \quad \beta = -1 \, .
\end{equation}
All in all, we find
\begin{equation}
    Z_4 \wedge W_4 = 8 I_\textbf{2} - I_\textbf{3} -13 I_\textbf{1} \, .
\end{equation}

\subsubsection*{Computation in $Sp(2)$}

In order to compute the branching rules for the anomaly polynomial under the inclusion $SU(2) \hookrightarrow Sp(2)$, we first need the anomaly polynomial of a $Sp(2)$ gauge theory. We can restrict to the fundamental and symmetric representation, which suffice to reproduce the fundamental and adjoint representations of $SU(2)$ under branching. We use the trace identities
\begin{eqaed}
    \text{tr}_\textbf{10}F^4 =& (4+8) \text{tr}F^4 +3(\text{tr}F^2)^2 = 12\text{tr}F^4 +3(\text{tr}F^2)^2 = 10 \, \frac{1}{2}(\text{tr}F^2)^2+ 3(\text{tr}F^2)^2 \, ,\\
    \text{tr}_\textbf{10}F^2 =& (4+2) \text{tr}F^2 = 6\text{tr}F^2
\end{eqaed}
for the symmetric rank-two representation. Since the anomaly polynomial for the fundamental representation is given by
\begin{equation}
    I_\textbf{4} = \left[\hat{A} \, \text{ch}_\textbf{4}(F)\right]\Bigg|_{8-\text{form}} = 4 \, I_\textbf{1} + \frac{1}{48} p_1\text{tr}F^2+\frac{1}{24}\text{tr}F^4 \, ,
\end{equation}
we obtain
\begin{equation}
    I_\textbf{10} = 10 \, I_\textbf{1} + \frac{1}{48} p_1\text{tr}_\textbf{10} F^2+\frac{1}{24}\text{tr}_\textbf{10}F^4 \, .
\end{equation}
Therefore, in terms of the fundamental traces,
\begin{equation}
    I_\textbf{10} = 10 I_\textbf{1} + \frac{1}{2} \, \text{tr}F^4 +\frac{1}{8}(\text{tr}F^2)^2 + p_1 \, \frac{1}{8}\text{tr}F^2 \, .
\end{equation}
From the above expression we observe that the sought-after coefficient of $\frac{1}{48} \, \text{tr} F^4$, taken modulo 12, does not depend on number of fermions in the $\mathbf{10}$, and equals twice the number $n_\mathbf{4}$ of $Sp(2)$ fundamentals (modulo 12). The branching rules
\begin{eqaed}
    \mathbf{4} & \to \mathbf{2} \oplus 2 \times \mathbf{1} \, , \\
    \mathbf{10} & \to \mathbf{3} \oplus 2 \times \mathbf{2} \oplus 3 \times \mathbf{1}
\end{eqaed}
then fix $n_\mathbf{4} = 10$, which finally gives $\mathcal{A}(1) = \frac{8}{12} = \frac{2}{3}$. This result matches \cite{Tachikawa:2024ucm}\footnote{In \cite{Tachikawa:2024ucm} there is an additional minus sign in the result due to a typo from eq. (3.26) to eq. (3.31), or (58)--(63) in the published version.}.

This concludes our review of the computation of the discrete topological term via subtraction of the two supersymmetric chiral spectra. In the next section, we compute the same term directly in the non-supersymmetric theory, using the chiral spectrum for \acp{NS5-brane} proposed in \cite{Basile:2023knk}.

\section{The non-supersymmetric NS5-brane}\label{sec:non-susy_NS5}

In this section we review the proposal of \cite{Basile:2023knk} for the chiral spectrum of the \ac{NS5-brane} in the heterotic non-supersymmetric theory. Inferring the spectrum of such an object is an involved problem. To begin with, in the absence of reliable string dualities\footnote{Some proposals of non-supersymmetric dualities were studied in \cite{Blum:1997cs, Blum:1997gw, Faraggi:2007tj}, and more recently in \cite{Basile:2022zee, Larotonda:2024thv} for the Sugimoto string \cite{Sugimoto:1999tx}.} one cannot identify the worldvolume degrees of freedom with some weakly coupled open-string sector. Moreover, without supersymmetry there are much fewer low-energy constraints on the spectrum. In the following we will retrace the anomaly inflow discussed in \cite{Basile:2023knk}, which leads to an educated guess for the chiral spectrum\footnote{See also \cite{Blaszczyk:2015zta} for a similar proposal.}. As emphasized in \cite{Basile:2023knk}, this cannot be seen as a proof, since different chiral degree of freedom can recombine to give the same anomaly. Nevertheless, we treat the proposed spectrum as a candidate to test in \cref{sec:discrete_top_term_NS5} and \cref{sec:YM}.

\subsection{Proposal for the chiral spectrum}\label{sec:proposed_spectrum}

With no further means to constrain the spectrum of the \ac{NS5-brane}, we have to resort to anomaly inflow. As we have reviewed in the preceding discussion, the chiral spectrum of the non-supersymmetric heterotic theory can be obtained as virtual difference of the (restrictions of the) two supersymmetric spectra,
\begin{eqaed}
    (\bf{16},\bf{16}) \ominus (\bf{128},\bf{1})\ominus(\bf{1},\bf{128})
    & = (\bf{120},\bf{1})\oplus (\bf{1},\bf{120}) \oplus (\bf{16},\bf{16}) \\
    & \ominus(\bf{128},\bf{1}) \ominus (\bf{1},\bf{128}) \ominus(\bf{120},\bf{1}) \ominus (\bf{1},\bf{120}) \, .
\end{eqaed}
Indeed, the first line represents the matter fields arising from the decomposition of the adjoint representation of $\mathfrak{so}(32)$ into representations of its $\mathfrak{so}(16) \oplus \mathfrak{so}(16)$ subalgebra. Similarly, the second line is the analogous in terms of the adjoint of $\mathfrak{e}_8 \oplus \mathfrak{e}_8$. Hence, the anomaly polynomials satisfy
\begin{eqaed}
    I_{12}^{SO(16)^2} = I_{12}^{\text{Spin}(32)/\mathbb{Z}_2}\big |_{SO(16)^2} - I_{12}^{E_8\times E_8} |_{SO(16)^2} \, ,
\end{eqaed}
and similarly for the Green-Schwarz terms
\begin{eqaed}
    Y_8^{SO(16)^2} = Y_8^{\text{Spin}(32)/\mathbb{Z}_2}|_{SO(16)^2} -  Y_8^{E_8\times E_8}|_{SO(16)^2}\,.
\end{eqaed}
Using these relations, one can perform an anomaly inflow with respect to the chiral spectrum in the background of a solitonic fivebrane, which is accessible in the effective field theory, and then probe the small limit in which the Yang-Mills configurations restore the full gauge group. In \cite{Basile:2023knk} it was shown that
\begin{eqaed}\label{eq:proposed_I8}
    Y_8^{SO(16)^2} & = \mathcal{I}^{(\bf{16},\bf{1})}_\text{D} + \mathcal{I}^{(\bf{1},\bf{16})}_\text{D} -\mathcal{I}_\text{SD} - 4\mathcal{I}^{(\mathbf{1},\mathbf{1})}_\text{D} \\
    &- \frac{1}{32}(c^{(1)}_{2}-c^{(2)}_{2})^2 -\frac{(c^{(1)}_{2}+c^{(2)}_{2}+3p_1)}{48} \, X_4^{SO(16)^2} \, ,
\end{eqaed}
where $\mathcal{I}^{\mathbf{R}}_\text{D}$ is the anomaly of fermion in the representation $\mathbf{R}$ of the gauge group, and $\mathcal{I}_\text{SD}$ the anomaly of a self-dual 2-form. Here, $c_2^{(i)}$ denote the second Chern classes pertaining to the two gauge group factors. This suggests that the chiral spectrum of the non-supersymmetric \ac{NS5-brane}, viewed as the small limit of the gauge soliton, is given by
\begin{itemize}
    \item four fermion singlets,
    \item a fermion in the $(\bf{16},\bf{1})\oplus(\bf{1},\bf{16})$ of $SO(16)^2$,
    \item a self-dual 2-form,
\end{itemize}
whose chiralities are determined by \cref{eq:proposed_I8}. The anomaly polynomial stemming from this chiral spectrum factorizes in the required form $I_8 = Z_4 \wedge W_4$. However, even without knowing the $SU(2)_\text{R}$ representations we can already see from \cref{eq:proposed_I8} that $W_4$ cannot be equal to that of \cref{eq:ZW_het}. Hence, we need to compute the global anomaly to check whether the resulting discrete topological term matches the computation of \cite{Tachikawa:2024ucm}. In order to do so, we first need to determine how $SU(2)_\text{R}$ refines the proposed spectrum.

\section{Discrete topological term from five-brane}\label{sec:discrete_top_term_NS5}

We now turn to the computation of the discrete topological term arising from the chiral \ac{NS5-brane} spectrum proposed in \cite{Basile:2023knk} and reviewed in \cref{sec:proposed_spectrum}. In particular, we want to assess its consistency with the results obtained in \cite{Tachikawa:2024ucm}, which we reviewed in \cref{sec:anomaly_inflow_review} and \cref{sec:global_anomaly_review}. Following the same approach, we extend the argument of \cite{Basile:2023knk} to include the $SU(2)_{\text{R}}$ normal bundle. Assuming that the representations comprise the trivial, fundamental and adjoint\footnote{A possible motivation for this ansatz stems from a putative duality with some D-brane spectrum \cite{Blum:1997cs, Blum:1997gw, Faraggi:2007tj, Basile:2022zee, Larotonda:2024thv}.}, the general spectrum takes the form
\begin{eqaed}\label{eq:SU2_general_anomaly_polynomial}
    \mathcal{I} \equiv n_\mathbf{1} \mathcal{I}_\mathbf{1} + n_\mathbf{2} \mathcal{I}_\mathbf{2} +n_\mathbf{3} \mathcal{I}_\mathbf{3} +n_{\text{SD}} \mathcal{I}_{\text{SD}} \, .
\end{eqaed}
We will now determine the $SU(2)_\text{R}$ (virtual) degeneracies by requiring consistency with anomaly inflow.

\subsection{Anomaly polynomial and factorization}\label{sec:anomaly_poly_NS5}
As we have discussed in \cref{sec:anomaly_inflow_review}, the \ac{NS5-brane} worldvolume effective action contains a localized Green-Schwarz term \cite{Green:1984bx,Sagnotti:1992qw, Mourad:1997uc, Imazato:2010qz}. Therefore, we must require that the anomaly polynomial $\mathcal{I}$ factorize with a factor of $Z_4 = \frac{p_1}{2} - 2 \, c_2$. To simplify the computation, we first choose the normal bundle to be trivial and account only for the gravitational anomaly. This means that the restriction $\mathcal{I}|_\text{grav}$ to the gravitational anomaly must be proportional to $p_1^2$. Each species contributes the anomaly polynomial $\mathcal{I}_\text{D}$ of an uncharged Dirac fermion. Therefore, the gravitational anomaly polynomial simplifies to
\begin{eqaed}\label{eq:SU2_general_anomaly_gravity_only}
     \mathcal{I}|_\text{grav} = (n_\mathbf{1}+2 n_\mathbf{2} + 3 n_\mathbf{3})\mathcal{I}^{(\mathbf{1}, \mathbf{1})}_\text{D} + n_{\text{SD}} \mathcal{I}_{\text{SD}} \, ,
\end{eqaed}
We now require that the gravitational anomaly spectrum reviewed in \cref{sec:proposed_spectrum} match this expression. The former simplifies to
\begin{eqaed}\label{eq:proposed_NS5_anomaly_gravity_only}
    32 \mathcal{I}^{(\mathbf{1}, \mathbf{1})}_\text{D}  -4\mathcal{I}^{(\mathbf{1}, \mathbf{1})}_\text{D}  - \mathcal{I}_\text{SD} = \frac{1}{32} \, p_1^2 \, . 
\end{eqaed}
Thus, matching \cref{eq:SU2_general_anomaly_gravity_only} and \cref{eq:proposed_NS5_anomaly_gravity_only} fixes
\begin{eqaed}\label{eq:matching_condition_1}
    n_\mathbf{1} & = 28 - 2n_\mathbf{2} - 3n_\mathbf{3} \, , \\
    n_\text{SD} & = - \, \frac{n_\mathbf{1} + 2 n_\mathbf{2} + 3 n_\mathbf{3}}{28} = -1 \, .
\end{eqaed}
The factorization must contain the combination $Z_4$, which implies that the polynomial should vanish upon replacing $p_1$ with $4c_2$. This requires
\begin{eqaed}\label{eq:matching_condition}
    n_\mathbf{2} = -2 - 8n_\mathbf{3} \, ,
\end{eqaed}
All in all, we find
\begin{eqaed}\label{eq:factorization_results}
    &n_\text{SD} = -1\,,\\
    &n_\mathbf{1} =32 + 13 n_\mathbf{3} \,,\\
    &n_\mathbf{2} = -2 - 8n_\mathbf{3}\,.
\end{eqaed}
At this stage, we have not fixed the adjoint degeneracy $n_\text{3}$ for $SU(2)_\text{R}$. We shall do so shortly, but for the time being we compute the global $SU(2)_\text{R}$ anomaly, which turns out not to depend on $n_\text{3}$ due to the modulo 12 condition.

\subsection{Global anomaly of the five-brane}\label{sec:global_anomaly_NS5}

As in \cref{sec:global_anomaly_review}, we can compute the global anomaly of the $SU(2)_\text{R}$ subgroup by extracting the coefficient $t$ in any $Sp(2)$ extension of the $SU(2)_\text{R}$ spectrum. As we have seen, the adjoint representation $\mathbf{10}$ of $Sp(2)$ does not contribute to $t$ modulo 12, since the term
\begin{eqaed}\label{eq:trF^4_term}
    \frac{24}{48} \, \text{tr} F^4
\end{eqaed}
comes with a factor of $24$. On the other hand, the fundamental representation $\mathbf{4}$, contributes $\frac{2}{48} \, \text{tr} F^4$. Therefore, one has
\begin{eqaed}\label{eq:t}
    t = 2 n_\mathbf{4} + 24 n_\mathbf{10} \equiv_{12} 2 n_\mathbf{4} \, .
\end{eqaed}
The branching rules under the inclusion $SU(2) \hookrightarrow Sp(2)$ yield
\begin{eqaed}
    n_\mathbf{4} \longrightarrow n_\mathbf{2}+2 n_\mathbf{1} \, .
\end{eqaed}
Together with \cref{eq:matching_condition}, this leads to
\begin{eqaed}\label{eq:result_for_t}
    t = 124 + 36 n_\mathbf{3} \equiv_12 4 \, ,
\end{eqaed}
giving the anomaly $\mathcal{A}(1) = \frac{4}{12} = \frac{1}{3}$ on the generator of $\pi_6(SU(2))$. As a $\mathbb{Z}_3$-valued topological term, our result is opposite to that of \cite{Tachikawa:2024ucm}, despite our efforts to consistently match all the conventions on chiralities, fluxes and anomalies. A possible explanation of this result is that the chiral spectrum proposed in \cite{Basile:2023knk} is not correct. Another possibility is that, since the computation of \cite{Tachikawa:2024ucm} contains an inflow from M-theory, the two computations differ by the physics captured by the M-theory inflow onto the E-string, analogously to how the Green-Schwarz term does not match the full \ac{NS5-brane} anomaly polynomial of the HO theory computed from its chiral spectrum \cite{Mourad:1997uc}, as shown in \cref{eq:Spin(32) inflow}. Indeed, for instance, the gravitational anomaly of a $\mathcal{N} = (1,0)$ tensor multiplet plus the free hypermultiplet of the E-string differs by the result of \cite{Ohmori:2014pca} by a factorized term proportional to $p_1^2$. It would be very interesting to clarify the origin of the discrepancy in the topological terms and settle this issue in future work. Another way to extract a discrete topological term is to include the $SU(2)_\text{R}$ representations directly in the inflow computation for the gauge soliton performed in \cite{Basile:2023knk}. Due to the standard embedding of the gauge and tangent bundles, the $SU(2)_\text{R}$ is identified with the instanton class of the gauge field profile. This determines the $SU(2)_\text{R}$ representations of the fermion zero modes as equal numbers of left-moving and right-moving doublets, packaged into different representations of the spacetime gauge group\footnote{We thank Y. Tachikawa for useful correspondence on this point.}. This suggests that the mixed anomaly mentioned in \cite{Tachikawa:2024ucm} could play a role in determining the discrete topological term computed by this worldvolume anomaly, which in any case differs from the other two anomaly polynomials by a factorized term as it should \cite{Basile:2023knk}. Resolving the ambiguity of which anomaly polynomial to use may require an understanding of a weakly coupled dual of the non-supersymmetric \ac{NS5-brane}, if any.

\section{Including Yang-Mills fields}\label{sec:YM}

In the preceding section we have seen that there is some freedom in the $SU(2)_\text{R}$ degeneracies of the \ac{NS5-brane} spectrum proposed in \cite{Basile:2023knk}, parametrized by the adjoint degeneracy $n_\text{3}$. Here we account for the gauge bundle and study the possible restrictions on $n_\mathbf{3}$. As we shall see, the complete anomaly inflow including both the gauge bundle and the $SU(2)_\text{R}$ bundle is consistent. We begin by observing that the local Green-Schwarz term in the $SO(16)^2$ is purely given by Chern classes of the gauge group,
\begin{eqaed}\label{eq:X8_so16xso16}
    Y_8^{SO(16)^2} = \frac{1}{24}\left(c_{2,\mathbf{16}}^{(1)\,2} + c_{2,\mathbf{16}}^{(2)\,2} + c_{2,\mathbf{16}}^{(1)}c_{2,\mathbf{16}}^{(2)} - 4 c_{4,\mathbf{16}}^{(1)} - 4 c_{4,\mathbf{16}}^{(2)}\right) ,
\end{eqaed}
while the Bianchi class $X_4$ is
\begin{eqaed}\label{eq:X4_so16xso16}
    X_4^{SO(16)^2} = \frac{1}{2}\left(p_1 +c_{2,\mathbf{16}}^{(1)}+c_{2,\mathbf{16}}^{(2)}\right) .
\end{eqaed}
In the decomposition of \cref{eq:proposed_I8}, we isolate the gauge contributions using
\begin{eqaed}
    \mathcal{I}^{\mathbf{16}}_\text{D} = \frac{c^{2}_{2}(F) - 2 c_{4}(F)}{12} + \frac{ c_{2}(F) \,  p_1}{24}  + 16 \, \mathcal{I}^{\mathbf{1}}_\text{D} \,.
\end{eqaed}
At this point, we include the $SU(2)_{\text{R}}$ representations. From \cref{eq:factorization_results} we infer that neither the self-dual field nor the $(\mathbf{16},\mathbf{1})\oplus(\mathbf{1},\mathbf{16})$ fermions can be packaged into a non-trivial $SU(2)_\text{R}$ representation, for otherwise the corresponding dimensions would be inconsistent with \cref{eq:factorization_results}. Therefore, the contribution of the four singlets is refined to
\begin{eqaed}
   -4\mathcal{I}^{(\mathbf{1}, \mathbf{1})}_\text{D} \longrightarrow 13 n_\mathbf{3}\, \mathcal{I}^{(\mathbf{1}, \mathbf{1})}_\text{D} - (2 + 8 n_\mathbf{3}) \mathcal{I}_\mathbf{2} + n_\mathbf{3}\mathcal{I}_\mathbf{3} \, ,
\end{eqaed}
while $X_4$ is refined to
\begin{eqaed}
    X_4^{SO(16)^2} \longrightarrow Z_4 = \frac{p_1}{2} - 2c_2(R) + \frac{c_{2}^{(1)} + c_{2}^{(2)}}{2} \,.
\end{eqaed}
When Yang-Mills fields are non-trivial, the structure of the factorization for $\mathcal{I} - Y_8$ is preserved, but the remainder is only consistent with \cref{eq:proposed_I8} for $n_\mathbf{3}=0$. To see this, turning on just the $SU(2)_\text{R}$ bundle the total anomaly polynomial $\mathcal{I}$ simplifies to
\begin{eqaed}
    \mathcal{I}|_{SU(2)_\text{R}} = (\mathcal{I}-Y_8^{SO(16)^2})\big|_{SU(2)_\text{R}} = \frac{p_1 - 4c_2}{2} \times \frac{3p_1 + 4c_2(1-4n_\mathbf{3})}{48} \, .
\end{eqaed}
Therefore, including the gauge bundle,
\begin{eqaed}\label{eq:SU2_lift}
    \mathcal{I} & = \frac{p_1 - 4c_2}{2} \times \frac{3p_1 + 4c_2(1-4n_\mathbf{3})}{48} + (\mathcal{I} - \mathcal{I}|_{SU(2)_\text{R}}) \\
    & = \frac{p_1 - 4c_2}{2} \times \frac{3p_1 + 4c_2(1-4n_\mathbf{3})}{48} + \left( \mathcal{I}_\text{D}^{(\mathbf{16}, \mathbf{1})} + \mathcal{I}_\text{D}^{(\mathbf{16}, \mathbf{1})} - 32 \, \mathcal{I}_\text{D}^{(\mathbf{1}, \mathbf{1})} \right) .
\end{eqaed}
Now, since the spacetime Green-Schwarz term of \cref{eq:X8_so16xso16} contains no Pontryagin classes, its refinement including the $SU(2)_\text{R}$ bundle cannot contain $c_2$ consistently with its ten-dimensional interpretation. Therefore, applying the decomposition of \cref{eq:proposed_I8} to the second term of \cref{eq:SU2_lift} leads to
\begin{eqaed}
    \mathcal{I} - Y_8^{SO(16)^2} & = \frac{p_1 - 4c_2}{2} \times \frac{3p_1 + 4c_2(1-4n_\mathbf{3})}{48} \\
    & - \, \frac{p_1^2 + (c_2^{(1)}+c_2^{(2)})^2}{32} + \frac{p_1 + c_2^{(1)} + c_2^{(2)}}{2} \times \frac{3p_1 + c_2^{(1)} + c_2^{(2)}}{48} \, .
\end{eqaed}
Setting $Z_4=0$, i.e. replacing $p_1$ with $4c_2 - c_2^{(1)} - c_2^{(2)}$, results in a term containing $c_2$. In more detail,
\begin{eqaed}
    \mathcal{I} - Y_8^{SO(16)^2} & = Z_4 \wedge \frac{3p_1 + c_2^{(1)} + c_2^{(2)} + 4(1-4n_\mathbf{3})c_2}{48} \\
    & + \frac{1}{96}(c_2^{(1)} + c_2^{(2)})(16 n_\mathbf{3}c_2-3c_2^{(1)} -3 c_2^{(2)}) \, .
\end{eqaed}
Such a residual dependence on the $SU(2)_\text{R}$ Chern class cannot be accounted for by the spacetime Green-Schwarz term, which was already subtracted. However, this dependence cancels precisely when $n_\mathbf{3}=0$. This is a rather natural assumption, although we cannot fully justify it; a stronger argument may involve a mixed anomaly along the lines discussed in \cite{Tachikawa:2024ucm}. Fixing $n_\mathbf{3} = 0$ gives $n_\mathbf{1} = 32$ and $n_\mathbf{2} = -2$. An intuitive way to state this result is that the $(\mathbf{16}, \mathbf{1}) \oplus (\mathbf{1}, \mathbf{16})$ fermions and the self-dual field are singlets of $SU(2)_\text{R}$, while the four gauge singlets repackage into two doublets of $SU(2)_\text{R}$.

\section{Bottom-up considerations}\label{sec:bottom-up}

In this section we consider more general bottom-up settings which can feature discrete topological terms induced by the Green-Schwarz mechanism. According to the discussion in \cite{Tachikawa:2023lwf, Tachikawa:2024ucm}, discrete topological terms in heterotic strings arise from the description of worldsheet theories in terms of \emph{topological modular forms}. Equivalence classes of worldsheet theories are encoded in elements of the abelian groups $\text{TMF}_\nu$, with $\nu = -22-d$ for the internal worldsheet theory of a heterotic vacuum with $d$ extended spacetime dimensions. This physical statement is grounded in the well-supported Stolz-Teichner conjecture \cite{Stolz:2011zj} and the cancellation of all global anomalies in supersymmetric vacua \cite{Tachikawa:2021mvw, Tachikawa:2021mby}. Since the discrete topological term, when well-defined, is an invariant of string bordism, it depends linearly on the classes $[M_d] \in \Omega^{\text{string}}_d$ of (compactified Euclidean) spacetime and $[T] \in \text{TMF}_{-d-22}$ of the internal worldsheet theory. Moreover, the discrete topological data is encoded in the subgroup $A_{-d-22} \subseteq \text{TMF}_{-d-22}$ defined by the kernel of the elliptic genus. Finally, we have the homomorphism
\begin{eqaed}
    & \Omega^{\text{string}}_d \to \text{TMF}_d \, , \\
    & [M] \mapsto [\sigma(M)]\,,
\end{eqaed}
which assigns the cobordism class of a representative string manifold to the deformation class of the corresponding (supersymmetric) sigma model. All in all, it is natural to propose that discrete topological terms be described by pairings
\begin{eqaed}
    A_d \times A_{-d-22} \to \mathbb{R}/\mathbb{Z} \, .
\end{eqaed}
This is indeed the case \cite{Tachikawa:2023lwf, Tachikawa:2024ucm}, and in the range of interest $A_d$ is non-trivial for spacetime dimensions $d=6,8,9,10$. In this paper, following \cite{Tachikawa:2024ucm}, so far we discussed the case $d=10$.

\subsection{Classification of topological terms}\label{sec:classification}

What are the other interesting options for discrete topological terms driven by the local Green-Schwarz mechanism? In order to have local anomalies, the spacetime dimension must be either $d=6$ or $8$. In both cases, the possible mediators of a local Green-Schwarz mechanism are an axion or a (local) two-form, up to dualities. 

By the completeness principle \cite{Polchinski:2003bq}, in the $d=6$ case, the former couples to an instanton or a three-brane, whereas the latter couples to a string, either electrically or magnetically. Assuming the presence of spacetime fermions to generate an anomaly in the first place, the relevant homotopy groups of the normal rotational symmetry are
\begin{eqaed}\label{eq:homotopy_6d}
    \pi_0(\text{Spin}(6)) \, , \quad \pi_4(\text{Spin}(2)) \, , \quad \pi_2(\text{Spin}(4))
\end{eqaed}
which all vanish. Thus, no anomalies of the type discussed in \cite{Tachikawa:2024ucm} and in the preceding sections arise.

For $d=8$, the axion can couple to either an instanton or a five-brane, whereas the two-form would couple to a string or a three-brane. In supergravity the anomalies automatically cancel, so we must consider non-supersymmetric gravity to find new possibilities. Out of the relevant homotopy groups encoding Witten anomalies for the normal bundle of worldvolume theories,
\begin{eqaed}\label{eq:homotopy_8d}
    \pi_0(\text{Spin}(8)) \, , \quad \pi_6(\text{Spin}(2)) \, , \quad \pi_2(\text{Spin}(6)) \, , \quad \pi_4(\text{Spin}(4)) \, ,
\end{eqaed}
only the last one is non-trivial. As in \cite{Tachikawa:2024ucm}, we shall focus on the $SU(2)_{\text{R}}$ subgroup, although it does not correspond to any R-symmetry. In summary, we are led to consider eight-dimensional gravity theories with a local Green-Schwarz mechanism mediated by a two-form. For simplicity, we shall restrict to a single two-form, although a generalization in the spirit of \cite{Sagnotti:1992qw} is in principle straightforward.

\subsection{Non-supersymmetric gravity in eight dimensions}\label{sec:8d}

In eight dimensions, the generator
\begin{eqaed}
\relax [SU(3)] \in \Omega^{\text{string}}_8 = \mathbb{Z}_2
\end{eqaed}
is represented by the $SU(3)$ group manifold. Since the inclusion $SU(2) \hookrightarrow SU(3)$ defines a fibration of $S^3$ over $S^5$, with $\pi_4(SU(3)) = 0$, we can apply the same strategy as \cite{Tachikawa:2024ucm} to characterize the associated discrete topological term.

The general structure of a factorized anomaly polynomial in $d=8$,
\begin{eqaed}\label{eq:8d_GS}
    \mathcal{I}_{10} = X_2 \wedge X_8 + X_4 \wedge X_6 \, ,
\end{eqaed}
shows that abelian gauge factors are necessary for the Green-Schwarz mechanism to be non-trivial. This means that the anomaly polynomial of the three-brane must be of the form
\begin{eqaed}
    \mathcal{I}_6^{\text{3-brane}} = X_6 + X_4 \wedge W_2 \, ,
\end{eqaed}
where $X_4$ is promoted to $Z_4$ upon turning on the normal $SU(2)_{\text{R}}$. However, turning off the gauge fields, these expressions must vanish for degree reasons. Therefore, lifting the anomaly theory from $SU(2)_{\text{R}}$ to $SU(3)$, as in \cite{Tachikawa:2024ucm}, would result in the unique term\footnote{In the case of $Sp(2)$, the (pseudo)reality of the fundamental representation and the eight-dimensional spinor representation imply that the index of the corresponding Dirac operator is even, and thus the minimal value for the integral of the degree-eight Chern character $\frac{1}{24} \, \text{tr} \, F^4$ over a closed manifold is two rather than unity \cite{Saito:2025idl}. In the case of $SU(3)$ this argument does not apply, and the relevant quantization of $t$, which is defined mod 2, is given by the degree-six Chern character.}
\begin{eqaed}
    \frac{it}{6} \tr F_{SU(3)}^3\,,
\end{eqaed}
corresponding to an anomaly $\mathcal{A}(1) = \frac{t}{2}$ for the generator $[SU(3)]$. Since the restriction to $SU(2)_{\text{R}}$ vanishes, the value of $t$ cannot be fixed purely by the $SU(2)_{\text{R}}$ representations of the worldvolume fermions. In order to remedy this, we now include a $U(1)$ gauge factor.

In the absence of gravitini, the chiral fermion content of the theory is parametrized by the degeneracy $n_q$ of each charge $q \in \mathbb{Z}$. The anomaly polynomial is
\begin{eqaed}\label{eq:8d_anomaly_polynomial}
    \mathcal{I}_{10} & = \sum_{q \in \mathbb{Z}} n_q \left(q^5 \, \frac{c_1(F)^5}{120} - \, q^3 \, \frac{p_1}{24} \, \frac{c_1(F)^3}{6} + q \, \frac{7p_1^2 - 4p_2}{5760} \, c_1(F) \right) ,
\end{eqaed}
where $c_1(F)$ denotes the first Chern class of the fundamental line bundle. For convenience, we define the moment sums $Q_k \equiv \sum_{q \in \mathbb{Z}} n_q \, q^k$. The above expression can then be factorized according to \cref{eq:8d_GS}, with
\begin{eqaed}\label{eq:8d_X4_X6}
    & X_4 \equiv \frac{p_1}{2} + \alpha \, c_1(F)^2 \, , \qquad X_6 \equiv \beta \, c_1(F)^3 - \, \frac{72\beta + Q_3}{144\alpha} \, p_1 \, c_1(F)
\end{eqaed}
and
\begin{eqaed}\label{eq:8d_X2_X8}
    & X_2 \equiv c_1(F) \, , \, X_8 \equiv - \, \frac{Q_1}{1440} \, p_2 + \frac{1440 \beta + 7 \alpha \, Q_1 +20 \, Q_3}{5760\alpha} \, p_1^2 + \frac{Q_5-120\alpha \beta}{120} \, c_1^4 \, .
\end{eqaed}
for any $\alpha, \beta$. In the above expressions we normalized the anomalous Bianchi identities by fixing the overall normalization of $X_4$ and $X_2$.

\subsubsection*{Three-brane probes}

Now we turn to the study of anomaly inflow onto the three-brane probes required by the completeness principle. Parametrizing by $d_q$ the charge degeneracies of the worldvolume fermions, the worldvolume anomaly polynomial is a function of the moment sums $D_k \equiv \sum_{q \in \mathbb{Z}} d_q \, q^k$. Denoting with $k$ the coefficient of the worldvolume Green-Schwarz term
\begin{eqaed}
    k \int B_2 \wedge c_1(F) \, ,
\end{eqaed}
one can solve the inflow equation
\begin{eqaed}\label{eq:3-brane_inflow}
    \mathcal{I}_6^{\text{3-brane}} = X_6 + k \, X_4 \wedge c_1
\end{eqaed}
for the parameters $\alpha, \beta$ in \cref{eq:8d_X4_X6}. At this point, we turn on the $SU(2)_{\text{R}}$ part of the normal bundle. The refined inflow must match the worldvolume anomaly polynomial with the right-hand side of \cref{eq:3-brane_inflow} modified according to
\begin{eqaed}\label{eq:3-brane_R_replacement}
    & X_4 \longrightarrow Z_4 \, , \\
    & X_6(p_1, c_1) \longrightarrow X_6(p_1 - 2c_2(R), c_1) \, .
\end{eqaed}
On the worldvolume of the three-brane, chiral fermions transform under $SU(2)_{\text{R}}$ representations, which we take to be singlets, doublets and triplets. The chiral spectrum is again parametrized by degeneracies $d_{q,\mathbf{1}}, d_{q,\mathbf{2}}, d_{q,\mathbf{3}}$. Thus, the anomaly polynomial contains moment sums $D_{k,\mathbf{1}}, D_{k,\mathbf{2}}, D_{k,\mathbf{3}}$, and takes the form
\begin{eqaed}\label{eq:3-brane_anomaly_polynomial_R}
    \mathcal{I}_6^{\text{3-brane}}  = & \left(D_{3,\mathbf{1}} + 2D_{3,\mathbf{2}} + 3D_{3,\mathbf{3}} \right) \frac{c_1(F)^3}{6} - \left(D_{1,\mathbf{1}} + 2D_{1,\mathbf{2}} + 3D_{1,\mathbf{3}} \right) \frac{p_1 c_1(F)}{24} \\
    & - \left(D_{1,\mathbf{2}} + 4D_{1,\mathbf{3}} \right) c_2(R) \, c_1(F) \, .
\end{eqaed}
The inflow in \cref{eq:3-brane_inflow} then fixes the Green-Schwarz factorization in \cref{eq:8d_X4_X6} according to\footnote{These coefficients may be subject to further integrality constraints, depending on the ultraviolet completion.}
\begin{eqaed}\label{eq:alpha_beta}
    \alpha =\frac{12 \left(D_{3,\mathbf{1}} + 2 D_{3,\mathbf{2}} + 3 D_{3,\mathbf{3}}\right) + Q_3}{6 \left(D_{1,\mathbf{1}} + 2 D_{1,\mathbf{2}} + 3 (8k + D_{1,\mathbf{3}})\right)} \, , \quad \beta = \frac{D_{3,\mathbf{1}} + 2D_{3,\mathbf{2}} + D_{3,\mathbf{3}}}{6} - k \, \alpha \, .
\end{eqaed}
However, \cref{eq:3-brane_inflow} also implies the condition
\begin{eqaed}\label{eq:3-brane_spectrum_constraint}
    D_{1,\mathbf{1}} + 14 D_{1,\mathbf{2}} + 51 D_{1,\mathbf{3}} = 12 k
\end{eqaed}
on the worldvolume spectrum.

We now lift \cref{eq:3-brane_anomaly_polynomial_R} to $SU(3) \times U(1)$ in order to compute the discrete topological term. We embed $SU(2)_{\text{R}}$ in $SU(3)$ in the standard fashion, so that the branching rules are
\begin{eqaed}\label{eq:SU3_branching}
    & \mathbf{8} \longrightarrow \mathbf{3} +2 \times \mathbf{2} +\mathbf{1} \, , \\
    & \mathbf{3} \longrightarrow \mathbf{2} +\mathbf{1} \, .
\end{eqaed}
Accordingly, the anomaly polynomials for each $SU(2)_{\text{R}}$ representation can be recast as
\begin{eqaed}\label{eq:SU3_poly_branching}
    \mathcal{I}^{SU(2)}_\mathbf{1} & = \mathcal{I}^{SU(3)}_\mathbf{1} \, , \\
    \mathcal{I}^{SU(2)}_\mathbf{2} & = \mathcal{I}^{SU(3)}_\mathbf{3} - \mathcal{I}^{SU(3)}_\mathbf{1} \, , \\
    \mathcal{I}^{SU(2)}_\mathbf{3} & = \mathcal{I}^{SU(3)}_\mathbf{8}- 2 \, \mathcal{I}^{SU(3)}_\mathbf{3} + \mathcal{I}^{SU(3)}_\mathbf{1} \, .
\end{eqaed}
Furthermore, the anomaly polynomials for the relevant representations of $SU(3)\times U(1)$ are
\begin{eqaed}\label{eq:SU3_anomaly_polynomials}
    \mathcal{I}^{SU(3)}_{q,\mathbf{8}}&= \frac{4}{3}q^3 c_1^3 - \frac{1}{3}q p_1 c_1 - c_2(R) c_1 \, , \\
    \mathcal{I}^{SU(3)}_{q,\mathbf{3}}&= \frac{1}{2}c_3 + \frac{1}{2}q^3 c_1^3 - \frac{1}{8}q p_1 c_1 - c_2(R) c_1 \, , \\
    \mathcal{I}^{SU(3)}_{q,\mathbf{1}}&= \frac{1}{6}q^3 c_1^3 - \frac{1}{24}q p_1 c_1  \, .
\end{eqaed}
The worldvolume anomaly polynomial of \cref{eq:3-brane_anomaly_polynomial_R} thus lifts to a combination of $SU(3) \times U(1)$ anomaly polynomials weighted by the original $d_{q,\mathbf{r}}$ degeneracies of $SU(2)_{\text{R}} \times U(1)$. One finds
\begin{eqaed}\label{eq:SU3_lift}
    \sum_{q \in \mathbb{Z}} \left( d_{q,\mathbf{3}} \, \mathcal{I}_{q, \mathbf{8}} + (d_{q,\mathbf{2}} - \, 2 d_{q,\mathbf{3}}) \, \mathcal{I}_{q, \mathbf{3}} + (d_{q,\mathbf{1}} - d_{q,\mathbf{2}} + d_{q,\mathbf{3}}) \, \mathcal{I}_{q, \mathbf{1}} \right) \, ,
\end{eqaed}
which simplifies to
\begin{eqaed}\label{eq:SU3_lift_simplified}
    & \frac{D_{0,\mathbf{2}} - 2 \, D_{0,\mathbf{3}}}{2} \, c_3(R) + \frac{D_{3,\mathbf{1}} + 2D_{3,\mathbf{2}} + 3D_{3,\mathbf{3}}}{6} \, c_1(F)^3 +\\
    & - \, \frac{D_{1,\mathbf{1}} + 2D_{1,\mathbf{2}} + 3D_{1,\mathbf{3}}}{24} \, p_1 \, c_1(F) - \left( D_{1,\mathbf{2}} + 4D_{1, \mathbf{3}}\right) c_2(R) \, c_1(F)\,.
\end{eqaed}
Therefore, we read off that the coefficient of $\frac{i}{6} \tr F^3_{SU(3)} = \frac{1}{2} \, c_3(R)$ is
\begin{eqaed}\label{eq:t_SU3}
    t = D_{0,\mathbf{2}} - 2 \, D_{0,\mathbf{3}} \equiv_2 D_{0,\mathbf{2}} \equiv \sum_{q \in \mathbb{Z}} d_{q,\mathbf{2}}\, ,
\end{eqaed}
which is nothing but the net number of chiral $SU(2)_{\text{R}}$ doublets in the worldvolume spectrum. Hence, the $\mathbb{Z}_2$-valued discrete topological term counts the parity of $SU(2)_\text{R}$ doublets. The connection between the spacetime and worldvolume spectrum is encoded in the fermion zero-modes in the background of the solitonic brane, as exploited in \cite{Basile:2023knk}. If the total gauge group of the spacetime theory supports such solitons and contains a $U(1)$ factor, we can use the branching rules
\begin{eqaed}
    S_+^{8d} & \to (S_+^{4d}, \mathbf{2}_\text{L}) \oplus (S_-^{4d}, \mathbf{2}_\text{R}) \, , \\
    S_-^{8d} & \to (S_-^{4d}, \mathbf{2}_\text{L}) \oplus (S_+^{4d}, \mathbf{2}_\text{R})
\end{eqaed}
for spinor representations of positive and negative chirality to obtain that the net number of $SU(2)_\text{R}$ worldvolume doublets is the net number of spacetime fermions, up to a sign which does not affect the discrete topological term.

\section{Conclusions} \label{sec:conclusions}

In this paper we approached the computation of the discrete topological term discovered in \cite{Tachikawa:2024ucm} directly from the non-supersymmetric $SO(16)^2$ heterotic string. Our method mirrors that of \cite{Tachikawa:2024ucm} to a large extent. The main difference is that we employed the proposed chiral spectrum for non-supersymmetric \acp{NS5-brane} discussed in \cite{Basile:2023knk}, extending it to include normal bundle representations. The discrete topological term turns out to be uniquely fixed by anomaly inflow from the anomaly polynomial we examined, as in \cite{Tachikawa:2024ucm}. However, we found the opposite result. Having attempted to track all possible sources of minus signs coming from chirality, flux and index-theoretic conventions, we are led to conclude that the source of the discrepancy is due to the physics captured by the two computations. In particular, the computation of \cite{Tachikawa:2024ucm} relies on an anomaly inflow from M-theory to obtain the anomaly polynomial of the HE \ac{NS5-brane}. As we have seen throughout our discussions, the difference between the index density of the chiral spectrum and an inflow polynomial can be physically meaningful, and this could potentially explain the difference between the two computations. More generally, any potentially viable anomaly polynomial would differ from ours by a factorized term, which can be compensated by a worldvolume Green-Schwarz term; another such polynomial including the $SU(2)_\text{R}$ bundle would arise by inflow from the gauge soliton via the standard embedding. Alternatively, the chiral spectrum proposed in \cite{Basile:2023knk} is not correct, and other proposals such as that in \cite{Blaszczyk:2015zta} should be considered. Indeed, the very notion of chiral spectrum for strongly coupled objects is subtle, especially without supersymmetry. This issue may be reflected in an ambiguity in the choice of anomaly polynomial that determines the topological term. If both non-zero values of the $\mathbb{Z}_3$-valued topological term were allowed and physically inequivalent, there should exist observables which are sensitive to its value. This would paint a heterotic counterpart of the picture discussed in \cite{Debray:2021vob}.

We also extended the method to more general settings from a bottom-up perspective. In the context of effective theories with a heterotic string realization, topological modular forms allow for a rather limited classification. In particular, in addition to the ten-dimensional case, purely gravitational discrete topological terms are non-trivial only for non-supersymmetric gravity in eight dimensions. In this setting, there is a $\mathbb{Z}_2$-valued discrete topological term of this type which is given by the parity of the number of chiral fermions. It would be interesting to investigate the landscape of heterotic constructions without spacetime supersymmetry in eight dimensions, combining the findings of \cite{Baykara:2024tjr} with the metastable flux compactifications of \cite{Basile:2018irz, Antonelli:2019nar}. It would be especially interesting if discrete topological term were constrained by quantum gravity consistency principles, which would produce novel swampland conditions. In the eight-dimensional case the full string bordism group is $\mathbb{Z}_2$, and thus no constraints from topological modular forms arise. This raises the intriguing possibility of finding further constraints by including spacetime gauge fields and consider twisted string bordism. This is expected to modify the allowed topological terms in a non-trivial fashion. The computation in \cite{Basile:2023knk} indicates that only $\mathbb{Z}_2$ factors appear in the degree-ten bordism group, which would have to be compared to the possible bilinear pairings given by (appropriately twisted) topological modular forms \cite{Lin:2024qqk}. A similar result already obtains for spin bordism of $B\text{Spin}(32)/\mathbb{Z}_2$ \cite{Kneissl:2024zox}, although these groups do not take into account the Bianchi identity.

At any rate, the power of anomalies and topology can be leveraged to learn more about non-perturbative aspects of string theory. Optimistically, this data could shed light on (S-)duality in the absence of spacetime supersymmetry as envisaged in \cite{Blum:1997cs, Blum:1997gw, Faraggi:2007tj, Basile:2022zee, Larotonda:2024thv}. Finding bottom-up constraints involving discrete data, such as the discrete topological terms we considered as well as quadratic refinements for self-dual fields \cite{Hsieh:2020jpj, Debray:2021vob, Dierigl:2022zll, Basile:2023zng, Dierigl:2025rfn}, would provide a novel robust tool to constrain quantum gravity effective theories admitting an ultraviolet completion, perhaps extending the results of \cite{Montero:2020icj, Bedroya:2021fbu} in this direction.

\section*{Acknowledgements}

The authors are grateful to Y. Tachikawa and H. Zhang for useful correspondence. I.B. thanks E. Kiritsis and H. Parra de Freitas for discussions. V.L. would like to thank C. Aoufia, M. Montero, S. Raucci and M. Tartaglia for insightful discussions and the Instituto de Física Teórica UAM-CSIC for the hospitality during the latest stage of this work. The work of I.B. is supported by the Origins Excellence Cluster and the German-Israel-Project (DIP) on Holography and the Swampland.

\printbibliography

\end{document}